\documentstyle[twocolumn,aps]{revtex}
\begin{document}
\draft
\title{
Metallic Spin Glass in Infinite Dimensions\cite{N}
}
\author{
O. Narikiyo
}
\address{
Department of Physics, 
Kyushu University, 
Fukuoka 810-8560, 
Japan
}
\date{
March, 2000
}
\maketitle
  Metallic spin glass is a new subject 
for theoretical studies, 
while experimentally such a state has been recognized 
for a long time. 
  Theoretical works by pioneers\cite{SRO,SG} have been done 
on the assumption of the presence of 
the metallic spin-glass state. 
  However, the metallic spin-glass state 
remains to be derived from a microscopic medel.\cite{RO} 
  In this Short Note 
we clarify the present status of the microscopic theory 
and develop it in comparison with the theory 
of the Mott transition in infinite dimensions.\cite{GKKR} 
  We will find a metal-insulator transition in the following 
and the transition is predominantly the Mott type 
but not the Anderson type. 

  We consider the random spin-fermion model 
in infinite dimensions introduced in ref.\ 2, 
\begin{equation}
H=-\sum_{ij\sigma}t_{ij}c^\dagger_{i\sigma}c_{j\sigma}
  +J_{\rm K}\sum_i S_i^z \sigma_i^z
  -\sum_{ij}J_{ij}S_i^z S_j^z,
\end{equation}
where $c_{i\sigma}$ represents the conduction electron 
with the spin $\sigma$, 
$\sigma_i^z$ the spin density of the conduction electron 
and $S_i^z$ the spin density of the localized spin at site $i$. 
  The conduction electrons hop on the Bethe lattice. 
  The density of states for bare conduction electrons 
is semicircular and the half-width is $2t$ 
where $t_{ij}=t/\sqrt{d}$ at a finite dimension $d$. 
  Although we did not introduce a randomness 
in the transfer integral $t_{ij}$, 
Gaussian randomness in $t_{ij}$ leads to a similar model to ours 
after averaging over the randomness.\cite{OR} 
  The exchange interaction, $J_{ij}$, between localized spins 
is a random variable obeying the Gaussian distribution 
as in the case of the mean field theory 
of the Ising spin glass.\cite{FH} 
  We assume for simplicity 
that the localized spins form a robust spin-glass state 
unaffected by the coupling, $J_{\rm K}$, to the conduction electrons. 
  Then we treat the localized spins as an environment 
which has a dynamics described by the mean field theory 
and drives the conduction electrons. 
  We are interested in the possibility of a metallic state 
so that we calculate the density of states of conduction electrons 
in the following. 
  For simplicity, we consider the half-filled case 
of the conduction electrons. 

  In infinite dimensions 
the momentum dependence of the self-energy for conduction electrons 
is absent\cite{GKKR} so that we only calculate the frequency dependence 
of the self-energy. 
  In the second order of the coupling, $J_{\rm K}$, 
the self-energy is given by 
\begin{equation}
\Sigma({\rm i}\omega_n)
=T J_{\rm K}^2 \sum_m \chi({\rm i}\Omega_m)
                      G_0({\rm i}\omega_n - {\rm i}\Omega_m),
\end{equation}
where 
$\chi({\rm i}\Omega_m)$ is the dynamical spin susceptibility 
of localized spins and $G_0({\rm i}\omega_n)$ the Green function 
of conduction electrons. 
  Here $\omega_n$ is a fermion frequency 
and $\Omega_m$ a boson frequency at a temperature $T$. 
  We consider the case of $T=0$ in the following. 
  Since it has been clarified in the study of the Mott transition\cite{GKKR} 
that the second-order perturbation 
gives a good interpolation between weak and strong coupling limits, 
we adopt the same approximation here and will see 
that it actually works well. 

  The spin susceptibility for localized spins 
has a static part $\chi_{\rm s}$ as 
\begin{equation}
\chi_{\rm s}={\Delta \over T} \delta_{m,0},
\end{equation}
in the spin-glass state of the mean field theory.\cite{FH} 
  Here $\Delta$ is the order parameter of the spin-glass state 
and the measure of broken ergodicity due to rugged energy landscape. 
  The mean field theory becomes exact in infinite dimensions. 
  We can introduce the dynamics of localized spins\cite{FH} 
and the retarded dynamical spin susceptibility has imaginary part 
proportional to $\omega^\nu$ 
where $\omega$ is a small real frequency and 
the exponent $\nu$ is about $1/4$ at $T=0$. 

  The Green function for conduction electrons is determined as 
\begin{equation}
G_0({\rm i}\omega_n)^{-1}={\rm i}\omega_n - t^2 G({\rm i}\omega_n),
\end{equation}
in the same manner as in the case of the Mott transition\cite{GKKR} 
where $G({\rm i}\omega_n)$ is the renormalized Green function 
for conduction electrons given by 
\begin{equation}
G({\rm i}\omega_n)^{-1}
=G_0({\rm i}\omega_n)^{-1} - \Sigma({\rm i}\omega_n).
\end{equation}
  In infinite dimensions 
$G({\rm i}\omega_n)$ plays the role of the dynamical mean field. 
  Since we study the half-filled case, 
the chemical potential, $\mu$, can be fixed as $\mu = 0$. 

  We calculate the density of states $\rho(\omega)$ 
of conduction electrons by $\rho(\omega)=-{\rm Im}G(\omega+{\rm i}0_+)/\pi$. 
  The density of states at the Fermi energy, $\rho(0)$, 
serves as the order parameter of the metal-insulator transition 
possible in our random spin-fermion model in infinite dimensions. 
  Namely, the transition is predominantly the Mott type. 
  Since our model is mapped onto a single-site model,\cite{GKKR} 
the localization character of the Anderson transition is apparently absent 
at the metal-insulator transition. 
  In the case of the Anderson transition $\rho(0)$ is uncritical. 
  On the other hand, $\rho(0)$ is critical 
at the Anderson-Mott transition in three dimensions.\cite{KB} 
  Although our solution is obtained in infinite dimensions, 
it might be relevant to the three dimensional system. 

  In our study the coupling $J_{\rm K}$ plays the same role 
as the local repulsion $U$ of the Hubbard model. 
  If we neglect the dynamics of $\chi(\omega)$, 
we obtain a metal-insulator transition of the Hubbard type.\cite{GKKR} 
  The transition obtained in ref.\ 3 is this type. 
  However, such a transition is an artifact due to insufficiency 
of the approximation. 
  The Mott transition should be ascribed to the quasiparticles 
of the Gutzwiller type\cite{GKKR} 
whose dynamics is drived by the dynamics of localized spins. 
  In our dynamical mean field theory 
both Hubbard type and Gutzwiller type characters are taken into account. 

  For simplicity, we approximate the dynamical susceptibility in eq.\ (2) 
by the static part in eq.\ (3). 
  This approximation favors the insulating state 
so that the actual $\lambda_{\rm c}$ is larger than the value 
obtained in the following. 
  The dynamics is taken into account as the origin of the width 
of the density of states of quasiparticles around the Fermi energy\cite{K} 
to be determined by the self-consistent procedure 
of the dynamical mean field theory. 

  A self-consistent numerical solution for $\rho(\omega)$ 
using the fast Fourier transform is given in Fig.\ 1 
where $J_{\rm K}\sqrt{\Delta}/t \equiv \lambda = 0.5$. 
  Here we have used the unit $t=1$. 
  The density of states roughly decomposed into three parts: 
the quasiparticle peak around the Fermi energy, $\omega=0$, 
and the upper and the lower Hubbard bands centered 
around $\omega=\pm J_{\rm K}\sqrt{\Delta}$. 
  The presence of the quasiparticle peak establishes 
the presence of a metallic state. 

  When we increase the value of the parameter $\lambda$, 
we have a metal-insulator transition at a critical value $\lambda_{\rm c}$ 
where the quasiparticle peak vanishes. 
  The transition point can be evaluated analytically 
in the same manner as in the case of the Hubbard model.\cite{RKZ} 
  In the metallic state near the transition point 
the self-energy at a low energy is estimated as 
\begin{equation}
\Sigma(\omega)
={J_{\rm K}^2\Delta \over 1+(4t^2/J_{\rm K}^2\Delta)}\cdot
 {\omega \over \omega^2 - \omega_0^2},
\end{equation}
where $\omega_0^2=t\ t^*/[1+(4t^2/J_{\rm K}^2\Delta)]$. 
  Here $t^*$ is the renormalized transfer integral for quasiparticles 
to be detremined self-consistently. 
  This self-energy serves as a good interpolation 
between weak and strong coupling limits. 
  If $\omega_0$ is finite, $\Sigma(\omega)\propto -\omega$ 
in the limit of $\omega \rightarrow 0$ 
so that the quasiparticle peak is obtained. 
  If $\omega_0$ vanishes, $\Sigma(\omega)\propto 1/\omega$ at small $\omega$ 
so that the quasiparticle peak vanishes and the Hubbard bands are obtained. 
  The self-consistently determined weight of the quasiparticle peak 
vanishes at $\lambda=\lambda_{\rm c}=1$ 
and this point corresponds to the metal-insulator transition. 

  In summary, we have combined the dynamical mean field theories 
of the Mott transition and the spin glass in infinite dimensions 
and derived a metallic spin-glass state from the random spin-fermion model. 
  Since our parameter $\lambda$ is proportional to $\sqrt{\Delta}$, 
which reflects the randomness in localized spin system, 
our theory has a relevance to the metal-insulator transition 
driven by randomness observed by experiments\cite{BK} 
where a spin-charge separation is established in the sense 
that the spin response exhibits an anomaly related to the existence 
of a spin-glass order, while the charge response is normal, 
in the metallic state. 

  We have many things undone in the present work. 
  They should be clarified in future studies. 
  For example, our theory is not fully self-consistent, 
since we neglect the modification of the spin susceptibility 
due to the coupling between conduction electrons and localized spins. 
  In order to understand the experiments, 
we have to develop a theory applicable to two or three dimensional case, 
while our present theory is formulated in infinite dimensions. 

  The author is grateful to Minokichi Sato, Hiroshi Kawai 
and Hitoshi Yamaoka for their support on establishing 
our computing environment. 


\end{document}